\documentclass[pra,reprint,superscriptaddress]{revtex4-1}
\usepackage{amsmath}
\usepackage{graphicx}
\usepackage{color}

\begin{document}

\title{In-plane electric fields and the $\nu=\frac{5}{2}$ fractional quantum Hall effect in a disk geometry}

\author{Anthony \surname{Tylan-Tyler}}
\email{atylan@pitt.edu} \affiliation{Department of Physics and Astronomy, University of Pittsburgh, Pittsburgh, Pennsylvania 15260, USA} 
\author{Yuli \surname{Lyanda-Geller}}
\email{yuli@purdue.edu}\affiliation{Department of Physics and Astronomy and Purdue Quantum Center, Purdue University, West Lafayette, Indiana 47907, USA}

\begin{abstract}
The $\nu=\frac{5}{2}$ fractional quantum Hall effect is of experimental and theoretical interest due to the possible non-Abelian statistics of the excitations in the electron liquid.
A small voltage difference across a sample applied in experiments to probe the system is often ignored in theoretical studies due to the Galilean invariance in the thermodynamic limit.
No experimental sample, however, is Galilean invariant.
In this work, we explore the effects of the probe electric fields in a disk geometry with finite thickness.
We find that weak probe fields enhance the Moore-Read Pfaffian state but sufficiently strong electric fields destroy the incompressible state.
In a disk geometry, the behavior of the system depends on the polarity of the applied radial field, which can potentially be observed in experiments using in a Corbino disk configuration.
Our simulation also shows that the application of such a field enhances the coherence length of quasiholes propagating through the edge channels. 

\end{abstract}

\maketitle

\textit{Introduction.}
The $\nu=\frac{5}{2}$ fractional quantum Hall effect (FQHE)  presents a unique experimental and theoretical challenge.
It is the only observed FQHE with an even denominator in electron single-layer systems \cite{ Review, 5/2Observation1,5/2Observation2,CsathyPRB}, making the non-interacting composite fermion \cite{Jain} and the related Laughlin pictures \cite{Laughlin} inapplicable.
As a result, more exotic models are considered, such as composite fermion pairing \cite{CooperPairing,StatPairing1,StatPairing2}.
Such pairing mechanisms lead to the Moore-Read Pfaffian and its particle-hole conjugate, the anti-Pfaffian \cite{MooreRead,ReadGreen,aPf1,aPf2}, which support non-Abelian quasihole/quasiparticle excitations \cite{MooreRead,nonabelian1,nonabelian2,nonabelian3,nonabelian4}, of interest for a topological quantum computer \cite{Freedman1998,NayakDasSarmaRPMQC, BonesteelSimonPRL, qubit1, qubit2}.
These non-Abelian quasihole excitations are responsible for the transport of charge through the system along the edge \cite{Edges}.

In order to probe these states, small voltages are applied in experiments.
Theoretically, in an infinite plane, small electric fields can be removed by a Galilean (Lorenz) transformation.
Experimental samples, particularly in the Corbino disk configuration, are not Galilean invariant and we thus expect that these fields will have measurable effects upon the ground state of the sample.
To consider the effects of these electric fields, we perform exact diagonalization calculations in a disk geometry with a finite sample thickness. 

We explore the phase diagram as the strength of the applied electric field, the Landau level (LL) mixing strength and interactions with the neutralizing background are varied.
We find regions where the total angular momentum makes the Moore-Read Pfaffian state plausible, and confirm its existence using overlap integrals with the known Moore-Read wavefunction  \cite{StatPairing1,StatPairing2}.
In a disk geometry, an electric field is applied in radial direction, similar to the experimental setup in Corbino disks.
Our simulations show that for electric fields in the $-\hat{r}$ direction, the overlap with the Moore-Read Pfaffian state first increases at fields well below the breakdown voltage before a precipitous drop.
In the $+\hat{r}$ direction, the overlap instead continuously decreases.
The disk geometry also provides an opportunity to study the edge states.
We find that the edge state coherence length increases five-fold with the inclusion of an in-plane electric field, regardless of direction \cite{Tylan-TylerLyanda-Geller}.
We suggest that exploration of the effects of an electric field on the excitation gap, especially in Corbino disk experiments \cite{SchmidtBennaceurBilodeauGervaisPfeifferWest}, may shed light on the nature of $\nu=5/2$ state and methods to control it.

\textit{Model.}
Our model system consists of a two dimensional electron gas (2DEG) of a thickness $w$ a distance $d$ above a disk of neutralizing background charge \cite{1/3Disk,5/2Disk,Tylan-TylerLyanda-Geller}.
This disk confines the electrons to a finite region, but also breaks particle-hole symmetry \cite{5/2Disk}.
A magnetic field $B$ is applied to the system.
Within the $\nu= 5/2$ state quantum Hall plateau, changing the magnetic field (with matching changes in the electron density) controls the strength of the LL mixing effects, $\kappa=\frac{e^2}{\varepsilon\ell_B}\frac{1}{\hbar\omega_c},$ where $\ell_B=\sqrt{\frac{\hbar c}{eB}}$ is the magnetic length and $\omega_c=\frac{eB}{mc}$ is the cyclotron frequency.
As we are interested in the half-filled first excited LL, we can fix the filling fraction $\nu$ by setting the radius of the disk $R=2\sqrt{N_e}\ell_B$ \cite{1/3Disk,5/2Disk,Tylan-TylerLyanda-Geller}, where $N_e$ is the number of electrons.
The first excited LL single particle eigenstates, in cylindrical coordinates, are 
\begin{eqnarray}
\label{state}
\psi_{1,m}(r,\theta,z)
&=&
\sqrt{\frac{1}{2\pi\ell_B^2(m+1)!}}\left(\frac{r}{\sqrt{2}\ell_B}\right)^me^{im\theta}e^{\frac{-r^2}{4\ell^2_B}}
\nonumber\\&&
\times
L_1^m\left(\frac{r^2}{2\ell_B^2}\right)\sqrt{\frac{2}{w}}\sin\frac{\pi z}{w}
\end{eqnarray}
where $m$ is the angular momentum and $w$ is the width of the rectangular potential well confining the 2DEG.

The interacting Hamiltonian is \cite{Tylan-TylerLyanda-Geller}
\begin{eqnarray}
\label{Hamiltonian}
H=H_{\text{1-Body}}+H_{\text{2-Body}}+H_{\text{3-Body}}.
\end{eqnarray}
The interactions of the electrons with the neutralizing background are contained in $H_{\text{1-Body}},$ while the electron-electron interactions are contained in the other two terms.
These are divided into two- and three-body interactions, which are calculated using a diagrammatic expansion of the Coulomb interaction, allowing quasiparticle excitations to higher and lower LLs \cite{Pseudo1,Pseudo2,Pseudo3,Pseudo4,Pseudo5,ED}.

This system was previously used to produce a phase diagram of the 5/2 FQHE \cite{5/2Disk,Tylan-TylerLyanda-Geller}, but here we are interested in the effects of an external, in-plane electric field applied to the sample.
We introduce this field by applying a constant voltage between a point contact at the center of the disk and the edge of the disk, which results in an additional one-body term in Eq. \ref{Hamiltonian}
\[
H_U=\sum_mU_mc^\dagger_mc_m,
\]
where $U$ is the strength of the applied field, $c_m^{(\dagger)}$ are the single electron annihilation (creation) operators, and matrix elements $U_m$ are
\begin{eqnarray}
\label{MatrixElements}
&U_m&=U\int_0^R\int_0^{2\pi}\int_0^wrdrd\theta dz r|\psi_{1,m}(r,\theta,z)|^2
\nonumber\\
&=&-\frac{\sqrt{2}U\ell_B}{(m+1)!}
\left[
\Gamma\left(j_m+2,\zeta\right)
-\Gamma\left(j_m+2,0\right)
\right.\nonumber\\&&\left.
-2(m+1)
\left(
\Gamma\left(j_m+1,\zeta\right)
-\Gamma\left(\j_m+1,0\right)
\right)
\right.\nonumber\\&&\left.
+(m+1)^2
\left(
\Gamma\left(j_m,\zeta\right)
-\Gamma\left(j_m,0\right)
\right)
\right]
\end{eqnarray}
where $j_m=m+3/2$, $\zeta=\frac{R^2}{2\ell_B^2}$, and $\Gamma(a,x)$ is the upper incomplete $\Gamma$ function.

\begin{figure}
\includegraphics{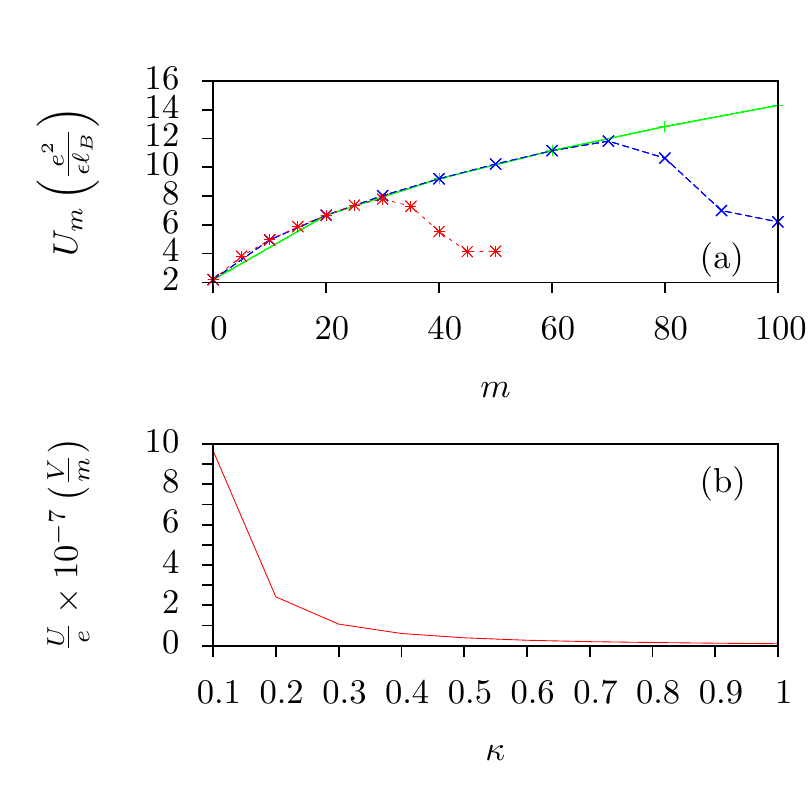}
\caption{
\label{Characteristics}
(a) Matrix elements of the external potential for systems with 25 (red), 50 (blue) and 100 (green) particles.
(b) As $\kappa$ is varied, the size of the electric potential for a fixed value for $U$ varies.
}
\end{figure}

In order to ensure that these terms are relevant in the thermodynamic limit, we look at their behavior as the system size grows.
The results are shown in Fig. \ref{Characteristics}a.
We see that there exist two clear regions which appear as a function of $m$.
The first region describes the bulk behavior, where $U_m\propto\sqrt{m}$.
As the edge is approached, however, there is a sharp drop in $U_m$ that results from the cut off in the field at the edge of the sample.
Thus, as in the bulk $U_m$ continues to grow and the edge region increases with sample size, we find that these fields will continue to have effects in the thermodynamic limit.

The in-plane electric fields of a quantum Hall state is limited by the breakdown voltage on the order of $10^3$V/cm \cite{WattsUsherMathewsZhuElliottHerrenden-HarkerMoorisSimmonsRitchie}.
As we measure energy in units of the Coulomb energy $\frac{e^2}{\varepsilon\ell_B}$, the field strength varies with $\kappa$ for fixed $U$ in Eq. \ref{MatrixElements}.
This variation is shown in Fig. \ref{Characteristics}b for the case $U=1\frac{e^2}{\varepsilon\ell_B^2}$.
The case $\kappa\rightarrow0$ is unphysical, but $\kappa>0.5,$ is the experimentally accessible regime.
Here we see that the applied field strength falls below the breakdown voltage for fairly large values of $U$ when compared to the LL mixing matrix elements. Making $U$ sufficiently small, we approach the strength of measurement fields, $\lesssim 1$V/cm.

\textit{Phase Diagram.}
As these effects persist in the thermodynamic limit and we can reasonably consider the effects of electric fields small enough to be within the measurement regime, we now explore the phase diagram as $d,$ $\kappa$ and $U$ are varied while $w$ is fixed at 1$\ell_B$.
To produce the phase diagram, an exact diagonalization calculation is performed for each $d$, $\kappa$ and $U$ in subspaces with fixed total angular momentum of the Hilbert space.
The state with the lowest energy in all such subspaces is then the ground state of the system and the state is labelled with that angular momentum \cite{5/2Disk,Tylan-TylerLyanda-Geller}.

Of primary interest to us here is the Moore-Read Pfaffian state.
This state appears with a a total angular momentum of $M_{\text{MR}}=N_e(2N_e-3)/2$.
Thus, using the total angular momentum of this state as well as overlap integrals, we can identify regions of our phase diagram which realize the Moore-Read Pfaffian.
In the current work we do not consider the anti-Pfaffian state as the zero field overlap is significantly smaller than that for the Moore-Read Pfaffian \cite{Tylan-TylerLyanda-Geller}.

\begin{figure}
\includegraphics{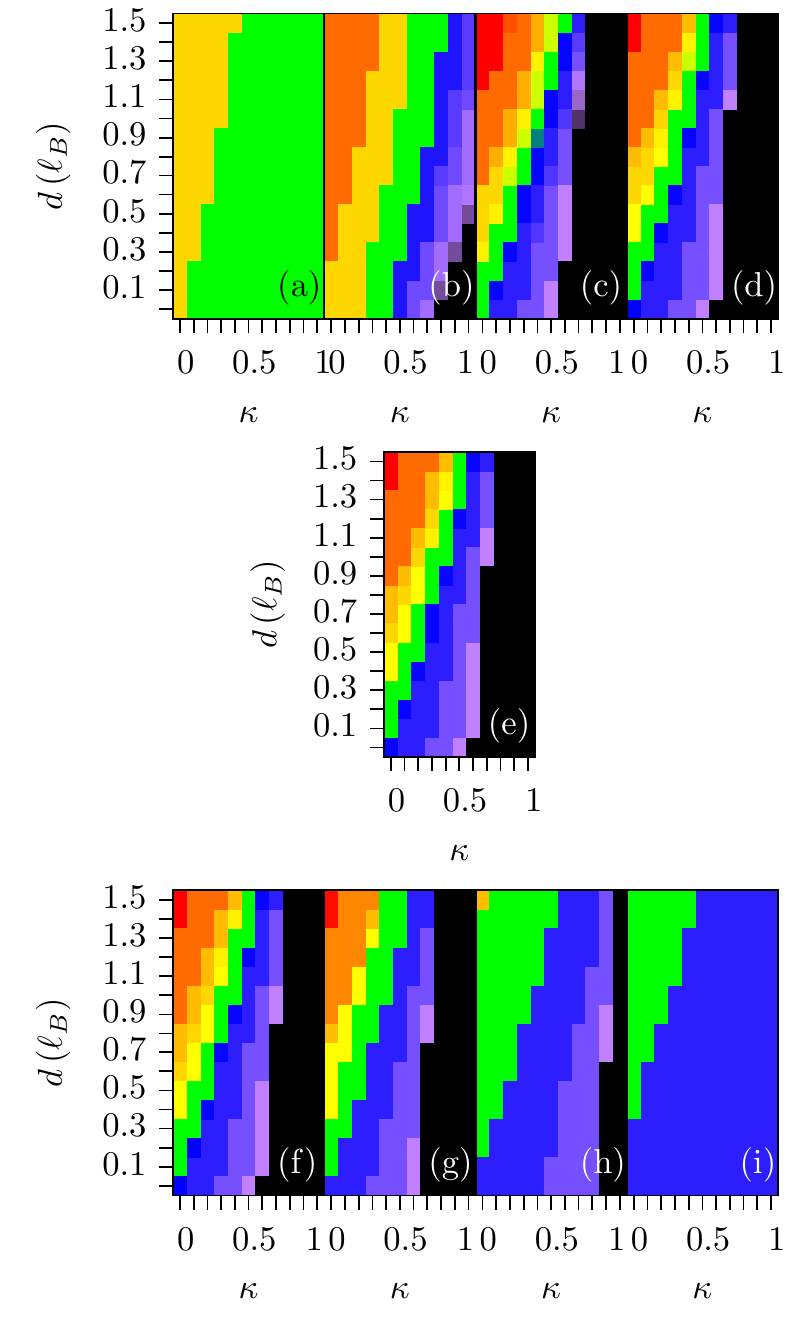}
\caption{
\label{Diagram}
The phase diagram as $U$, $d$, and $\kappa$ are varied for 10 electrons in 18 single particle states confined by a well of width $w=1\ell_B$ parallel to the magnetic field.
The green region in each figure has total angular momentum $M=85$, the total angular momentum of the Moore-Read Pfaffian for 10 particles.
In the top row, $|U|$ is decreasing, with (a) $U=0.1$, (b) $U=0.01$, (c) $U=0.001$, (d) $U=0.0001$.
The middle row is a reference case when (e) $U=0$, taken from Ref. \onlinecite{Tylan-TylerLyanda-Geller}.
In the bottom row, $|U|$ is increasing, with (f) $U=-0.0001$, (g) $U=-0.001$, (h) $U=-0.01$, (i) $U=-0.1$.
}
\end{figure}

We apply this process to 10 electrons in 18 single particle states.
By limiting the system to 18 states, an artificial hard wall potential is produced which limits edge reconstruction \cite{1/3Disk}, allowing us to access the bulk behavior.
The results are shown in Fig. \ref{Diagram}, with the field strength decreasing going right to left.
The case $U=0$, taken from Ref. \onlinecite{Tylan-TylerLyanda-Geller}, is shown in Fig. \ref{Diagram}e for comparison.

At first glance, we see that for the strongest fields, $U=\pm0.1\frac{e^2}{\varepsilon\ell_B},\pm0.01\frac{e^2}{\varepsilon\ell_B}$, that the green $M=85$ region associated with the Pfaffian has greatly expanded compared to $U=0$ case.
However, while the $M=85$ region has grown substantially, the overlap data shows that the Moore-Read Pfaffian has been destroyed, with the overlap $\sim0$ in the positive $U$ case and $\sim0.02$ in the negative $U$ case, compared to $\sim0.7$ at $U=0$.

The results in Fig. \ref{Characteristics}a for a strong positive $U$ can be understood from the minima of the effective potential.
The center of the disk and the edge regions have lower potentials than the central band of the disk.
Thus, for large $U$, the electric potential splits the 2DEG into two parts, isolated at the center and the edge of the disk respectively, forming a stripe state with total angular momentum $M=85$.

For strongly negative $U$, the same edge effect, which isolated a band of the 2DEG at the edge at $U>0$, now forms a strong barrier, and the 2DEG is isolated to a band between the center and edge of the disk.
Unlike the positive $U$ case, this state is not well described by a stripe state as no single Fock state dominates.

Turning now to the weak electric field regime, $|U|\leq0.001\frac{e^2}{\varepsilon\ell_B},$ we see that the behavior of the $M=85$ region is dependent upon the direction of the applied field.
For positive $U$, we see that the overlap increases with increasing field strength, to a maximum of $\sim0.8$ before falling off rapidly.
In contrast, for negative $U$, the overlap only falls off with increasing field strength.

This difference arises from how the fields act on the sample.
While increasingly positive $U$ leads to a depopulation of a central band of the disk, the competing effects of the edge and the core attractive potentials lead to a more homogeneous distribution, as expected from the Moore-Read state, until the breakdown voltage is reached and the state is destroyed.
In the case of an increasingly negative $U$, this balancing is absent and the electrons only pool increasingly in a limited region of the disk, thus depopulating the edge and core regions.
As a result of this depopulation, the negative fields lead to a decreasing overlap with the Moore-Read Pfaffian.

From the phase diagram, we see that as the fields increase in strength, the Pfaffian state is destabilized, with the overlap integral tending towards 0.
This behavior is expected from the observation of the breakdown voltage in experiment.
More interesting, we see that weak fields can lead to an increase in the overlap with the Moore-Read Pfaffian, increasing from a maximum of $\sim0.7$ with no field to $\sim0.8$ in the case of positive $U$, before the state is destroyed by the breakdown voltage.

\textit{Edge States.}
We now turn to the effects of the electric field on the edge states.
So far we used 10 electrons in 18 states to suppress potential edge reconstruction, allowing bulk effects to dominate \cite{1/3Disk,5/2Disk,Tylan-TylerLyanda-Geller}.
Now, we allow these edge effects to be prominent so that we may study the edge behavior of the system.
This is done by expanding the single-particle Hilbert space to 22 states, creating a softer edge potential.
We then find the wavefunctions and calculate the overlap integral with wavefunctions of the edge modes of the Moore-Read Pfaffian.
We find that the overlaps are $>0.1$. However, once we apply the electric field, the overlap of some of the edge modes fall below $0.1$.
In contrast, we find that the application of the electric field increases the overlap with the ground state, regardless of the direction of the applied field. This difference in behavior when compared to the bulk is the result of the applied negative $U$ potential acting as a confining potential, suppressing edge reconstruction effects.

\begin{figure}
\includegraphics{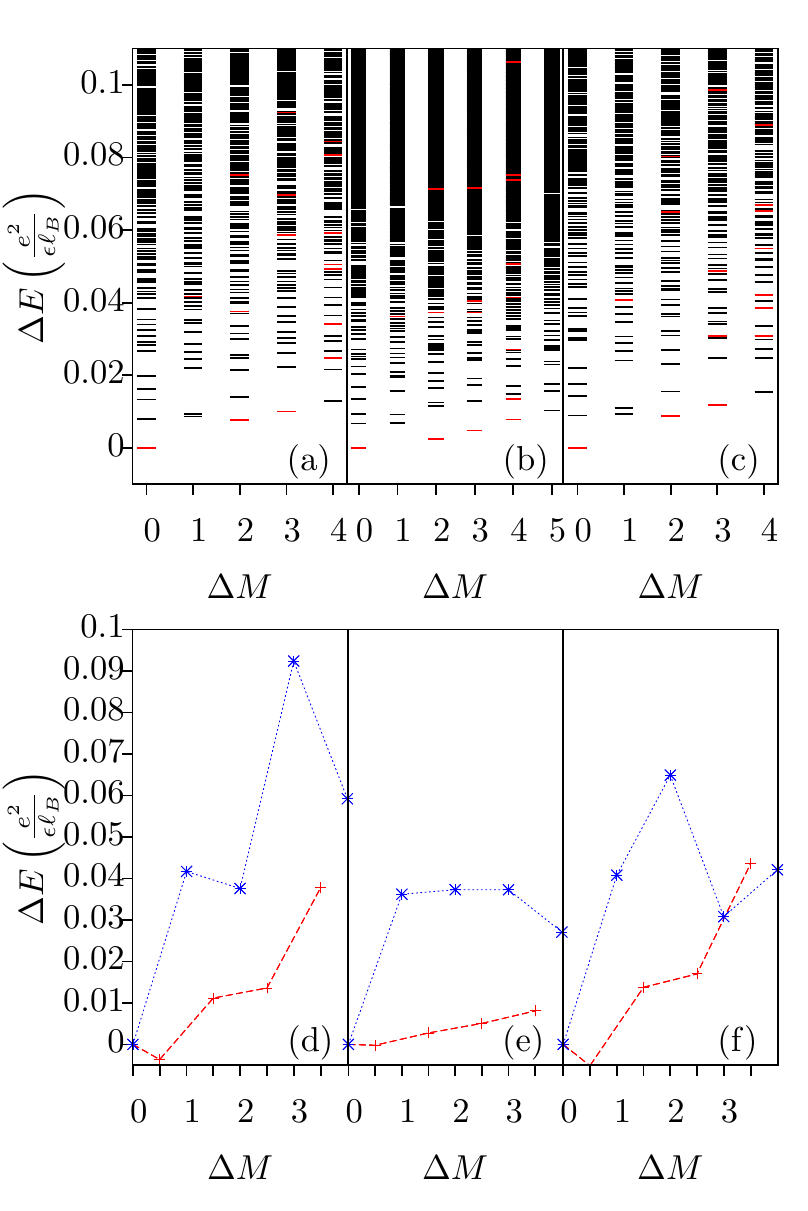}
\caption{
\label{Edges}
The energy spectra when $d=1\ell_B$, $\kappa=0.3$ and (a) $U=0.0001$, (b) $U=0$, (c) $U=-0.0001$, with the edge modes highlighted in red.
 (d) the dispersion relations $U=0.0001$, (e) $U=0$, (f) $U=-0.0001$.
An energy shift causes the Fermi modes to have a larger velocity and leads to increase of the coherence length of edge quasiparticles  compared to the $U=0$ case.
}
\end{figure}

The spectra for 10 electrons in 22 states are shown in Figs. \ref{Edges}a,b,c when $d=1\ell_B$ and $\kappa=0.3$.
For reference, the edge spectrum from Ref. \onlinecite{Tylan-TylerLyanda-Geller} is shown in Fig. \ref{Edges}b.
Comparing this to the spectra shown in Figs. \ref{Edges}a,c we see the same general features emerge: the Bose modes, carrying charge, are well mixed with the bulk, while the Fermi edge modes, carrying the quasiparticle statistics, are generally separated from the bulk.

Where the spectra differ from the $U=0$ case most significantly is the energy spacing.
The electric field increases the excitation energy from the ground state to the edge modes.
This increase results from the persistent current along the edge of the disk caused by an electric field.  As the Bose and Fermi modes carry components of the quasiholes along the edge, their energy also increases.

The effects of this acceleration are seen from comparison of Figs. \ref{Edges}d,f to Fig. \ref{Edges}e, reproduced here from Ref. \onlinecite{Tylan-TylerLyanda-Geller}.
The increase in energy is shown by the larger slope in the dispersion of the Fermi and Bose modes when compared to the $U=0$ case.
Also from the dispersion relations, we find evidence of the Bose modes distinctly carrying charge, as the behavior of the Bose mode dispersion changes with the polarity of $U$ while the behavior of Fermi modes is unaffected.

From the edge dispersion, we can also calculate the coherence length \cite{CoherenceLength,5/2Disk,Tylan-TylerLyanda-Geller}
For the $U=0$ case, this gives us a coherence length of $L_\phi\simeq2.82\mu$m \cite{Tylan-TylerLyanda-Geller}.
With the application of the electric field, we find that the coherence length increases to $L_\phi\simeq9.7\mu$m and $\simeq10.7\mu$m for positive and negative $U$ respectively.
Thus, the application of the electric field and the resulting edge currents substantially improve the edge coherence lengths of the quasiholes.

\textit{Conclusion.}
Applying an external electric field with a magnitude exceeding a nominal probe field to a quantum Hall fluid has many beneficial effects.
When edge effects are suppressed the behavior is dependent upon the polarity of the field, with negative $U$ decreasing the overlap and positive $U$ raising the overlap with Moore-read state long before the breakdown voltage is reached.
At high voltage, the Moore-Read state collapses, with the strongly positive $U$ case going to a stripe state with the 2DEG in isolated bands 
at the center and very edge of the disk, while the strongly negative $U$ case sees the 2DEG collapse between the edge and center of the disk.
These direction-dependent changes from the $U=0$ case are the result of the form of the matrix elements of radial electric field potential.

For experimentally accessible edge states, small applied fields strengthen the Moore-Read state.
With the relaxation of the hard wall conditions, the case of weak negative $U$ shows an increase in the ground state overlap as a result of an additional confinement of electrons to the disk, similar to the effects of LL mixing as the 2DEG is moved further from the neutralizing background.
Weak fields also lead to significant improvement of the edge coherence length.
Thus, the in-plane fields  can be used to fine-tune the ground state and improve the edge properties of the non-Abelian states.

It is therefore of interest to explore the effects of increasing voltages used to measure the quantum Hall signals.
Similar to recent experiments applying hydrostatic pressure to the sample \cite{Csathy_p}, this offers  means of dynamically tuning the ground state of the half-filled first excited LL between the incompressible Moore-Read Pfaffian and a compressible stripe state.
A Corbino disk configuration would be of particular interest as the effects of the orientation of the electric field can be probed.

\begin{acknowledgments}
This work was supported by the U.S. Department of Energy, Office of Basic Energy Sciences, Division of Materials Sciences and Engineering under Award {DE-SC0010544}.
\end{acknowledgments}

\end{document}